# Effects of high energy electron irradiation on quantum emitters in hexagonal boron nitride

Hanh Ngoc My Duong[1,†], Minh Anh Phan Nguyen[1,†], Mehran Kianinia[1], Hiroshi Abe[2], Takeshi Ohshima[2], Kenji Watanabe[3], Takashi Taniguchi[3], James H. Edgar[4], Igor Aharonovich[1,*], and Milos Toth[1,*]

1. School of Mathematical and Physical Sciences, University of Technology Sydney, Ultimo, New South Wales, 2007, Australia
2. National Institutes for Quantum and Radiological Science and Technology, 1233 Watanuki, Takasaki, Gunma 370-1292, Japan
3. National Institute for Materials Science, 1-1 Namiki Tsukuba Ibaraki 305-0044 Japan
4. Department of Chemical Engineering, Durland Hall, Kansas State University, Manhattan, KS 66506, USA

† These authors contributed equally to this work
*igor.aharonovich@uts.edu.au, *milos.toth@uts.edu.au

ABSTRACT:
Hexagonal Boron Nitride (hBN) mono and multilayers are promising hosts for room temperature single photon emitters (SPEs). In this work we explore high energy (~ MeV) electron irradiation as a means to generate stable SPEs in hBN. We investigate four types of exfoliated hBN flakes - namely, high purity multilayers, isotopically pure hBN, carbon rich hBN multilayers and monolayered material - and find that electron irradiation increases emitter concentrations dramatically in all samples. Furthermore, the engineered emitters are located throughout hBN flakes (not only at flake edges or grain boundaries), and do not require activation by high temperature annealing of the host material after electron exposure. Our results provide important insights into controlled formation of hBN SPEs and may aid in identification of their crystallographic origin.

INTRODUCTION:
Solid state materials capable of hosting optically active single photon emitters (SPEs) are highly sought after for applications in quantum nanophotonics and quantum information processing[1-3]. To this extent, defects in diamond, silicon carbide and gallium nitride have been studied as promising candidates for SPEs[4-8]. The main advantage of this class of SPEs is their optically-stable room-temperature operation. Recently, hexagonal boron nitride (hBN) has been identified as a host of ultra-bright, room-temperature SPEs[9-19]. A significant advantage of hBN is its layered nature, which enables convenient preparation of very thin single crystal layers (~ < 50 nm) and monolayers. hBN is also a wide bandgap material (~ 6 eV), which is advantageous for creating stable, isolated, deep-trap defect states.

SPEs in hBN have so far been observed in a range of materials, including bulk crystals[20-21], mechanically-exfoliated multilayers[12, 14], liquid-exfoliated flakes[10], nanotubes and other nanostructures[22-23]. hBN samples often require high-temperature post-growth annealing in order for stable emitters to be observed, and it is not clear whether the annealing treatments create emitters or activate pre-existing defects. Recent efforts showed that plasma treatment[24], ion implantation[12] or strain[25] can potentially be used to create/activate emitters. However, emitters in hBN possess a broad range of photophysical properties, and the atomic structure of the defects is a matter of debate[26-29]. In addition, emitters in large-area mechanically-exfoliated flakes are often located at grain boundaries or interfaces and emitters in large-area monolayers are sparse[12, 22, 24].

In this work, we investigate the effects of high energy (2 MeV) electron irradiation[30-31] on a range of hBN samples (summarized in table 1). The electron exposure generates SPEs, via either the generation of new defects or optical activation of existing defects, and the generated emitters were characterized using photoluminescence (PL), time-resolved measurements and photon correlation analysis. Our results suggest that the electron irradiation process can be used to engineer optically-active defects without the need for high temperature annealing and the emitters can be located within the hBN flakes rather than just flake edges or grain boundaries. Finally we show that upon MeV electron irradiation, single emitters can be created in large area hBN monolayers - a task that has previously been elusive. Our results provide important insights into controlled generation of SPEs in hBN and will aid future studies of the atomic structure of the defects.

EXPERIMENTAL SECTION:
**hBN Sample Preparation:**
Four different types of samples were prepared; multilayer hBN flakes from high purity hBN (purchased from HQ Graphene), multilayer flakes from carbon-enriched hBN, multilayer flakes made from isotopically-enriched Boron-10 (99.2% $^{10}$B) and hBN monolayers (purchased from Graphene Supermarket) that were grown by chemical vapour deposition (CVD).

High-purity flakes purchased from HQ graphene were synthesised by reacting fused boric acid with urea to form $B_2O_3.xNH_3$ and then heated further in ammonia to 900 °C. The product was further heated in a nitrogen atmosphere at 1500 °C, allowing it to crystallize on a substrate surface.

The isotopically-enriched $^{10}$B flakes were grown with a high purity Ni-Cr metal mix with enriched $^{10}$B powder in an alumina boat at 1550 °C for 24 hrs under an atmosphere of forming gas and nitrogen in a 1:4 ratio. It was subsequently cooled at a rate of 1°C/hr until 1500 °C, whereby it was quenched to room temperature. This sample is referred to as B10 in the manuscript.

The multilayer samples were mechanically exfoliated with scotch tape onto silicon substrates and baked at 450 ºC in air to remove residual adhesives and other unwanted carbonaceous material.

Monolayer hBN grown on Cu foil by CVD was cut into small squares and transferred onto Si substrates. The foils were then coated with PMMA to provide a platform to facilitate adhesion of the sample. The PMMA film containing the monolayer was then undercut by chemical etching of the Cu substrate with iron nitrate, leaving the polymer layer floating above. The film was transferred to a silicon substrate and heated to 120 °C for 20 minutes to promote adhesion, before being washed with acetone to dissociate the polymer.

**Electron irradiation**

Electron irradiation was performed with a Cockcroft-Walton 2 MV 60 kW electron accelerator. The samples were wrapped in Al foil and placed on a water-cooled Cu plate for the irradiation process. The cooling ensured that the sample temperature was kept fixed at approximately room temperature (~25 °C) during electron processing. Irradiation was performed at atmospheric conditions. The as-prepared samples were irradiated at 2 MeV with a fluence of 1 x $10^{15}$/cm$^2$ for all hBN samples.

**Optical characterisation:**

The irradiated samples were characterized using a home-built confocal PL setup with a 532 nm excitation laser. Details of the apparatus have been provided elsewhere[10]. Cryogenic measurements were performed at 10 K using a Janis ST500 cryostat system integrated with a high numerical aperture objective and 3-axis piezo sample positioning system.

RESULTS AND DISCUSSION:

Figure 1 details PL characterisation of emitters in a high-purity, single-crystal, multilayer flake of hBN. Figure 1a shows a confocal PL scan of a flake after electron irradiation. This flake did not contain any single luminescent centers before electron exposure. After electron irradiation, there was a high density of single luminescent emitters within the flake. A representative spectrum is shown in Figure 1b, with a sharp zero-phonon line (ZPL) at ~ 590 nm and distinct phonon-sideband (PSB), red-shifted by ~ 54 nm. The quantum nature of the emission is confirmed by measuring the second-order autocorrelation function, $g^{(2)}(\tau)$, as shown in the inset. The dip at zero delay time, $g^{(2)}(0)$ ~ 0.26, is indicative of a single-photon emitter.

Figure 1c shows a confocal PL scan of (a different region of) the same irradiated flake after an additional annealing step performed at 750 °C under 1 torr of argon for 30 minutes. Figure 1d shows a representative spectrum of an emitter obtained after annealing (performed after electron irradiation). The quantum nature of the defect is once again confirmed by the $g^{(2)}(\tau)$ function (Figure 1d inset). Overall, the electron irradiation process created a distribution of ZPL energies, with preferential creation of luminescent defects emitting at ~ 600 nm when excited with a 532 nm laser. There are no substantial, systematic differences between the properties of emitters created by electron irradiation only, annealing only, or a combination of irradiation and annealing. The annealing treatment can both create and destroy emitters, as has been reported previously[31].

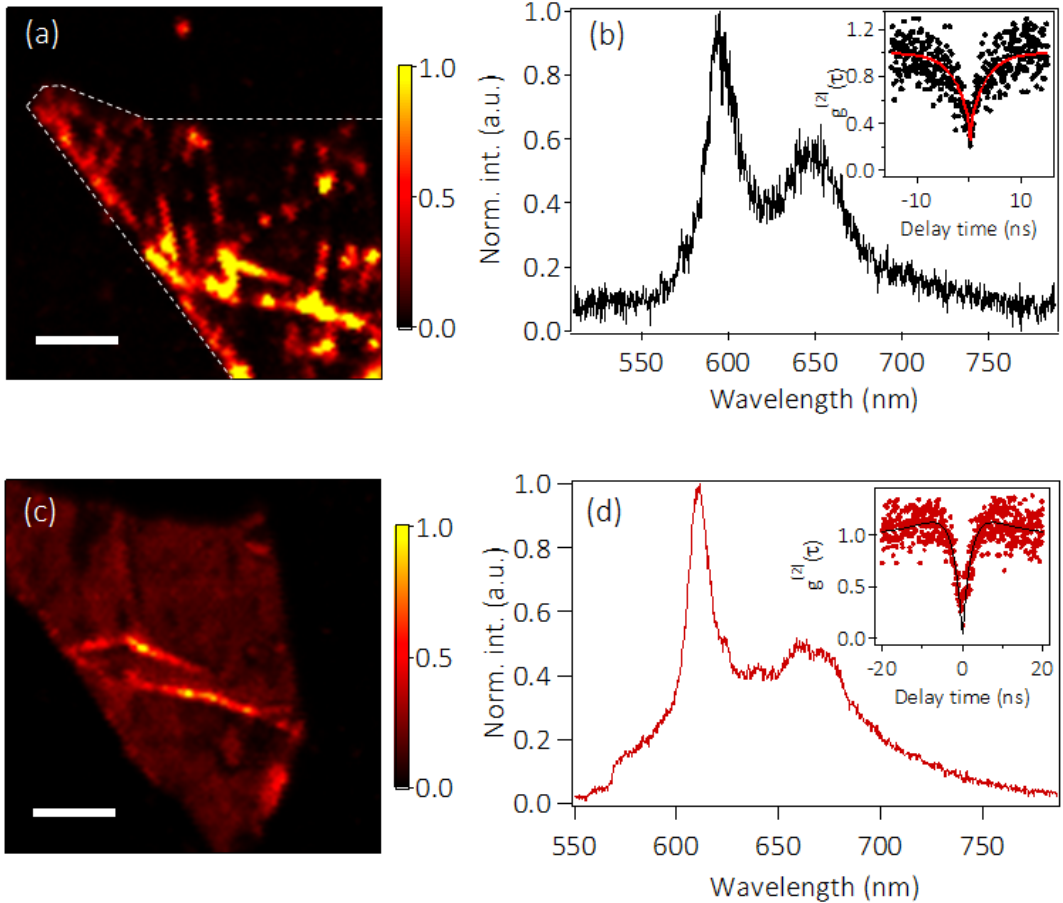

*Figure 1. Effects of electron irradiation and on high purity, single crystal, multilayer flakes of hBN. (a) Confocal PL map of an irradiated flake. (b) Typical PL spectrum of an emitter formed in irradiated flakes. The inset shows the corresponding second-order autocorrelation function. (c) PL map of a different region of the flake obtained after high temperature annealing. (d) Typical PL spectrum of an emitter obtained after electron irradiation and thermal annealing. The inset shows the second-order autocorrelation function. Both spectra were collected at 0.3 mW for 10 seconds at room temperature. All scale bars are 5 μm.*

To date, attempts to deterministically engineer single emitters in hBN have resulted with emitters located predominantly at edges and grain boundaries[22, 25, 31]. In contrast, the MeV electron irradiation can produce emitters both at edges/boundaries, and in the central, flat areas of flakes. Figure 2a shows an optical image of a large multilayered hBN flake that was subject to MeV electron irradiation. The thickness of the flake is 28.4 ±0.4 nm as confirmed by atomic force microscopy (AFM) (see inset). Confocal PL maps of the same flake reveal emitters localized within the flake (white circles in Figure 2b). No grain boundaries or edges are visible near the emitters. The morphology of the flake is further confirmed by a high resolution AFM scan, as shown in figure 2c. Figure 2d shows a representative spectrum from one of the formed emitters, labeled '1' on the confocal map. The inset is a corresponding second order autocorrelation function, confirming the quantum nature of the emitter ($g^{(2)}(0)$ = 0.12). We note that a subsequent annealing treatment destroyed all of the emitters shown in Figure 2, and generated new emitters located at flake edges and grain boundaries.

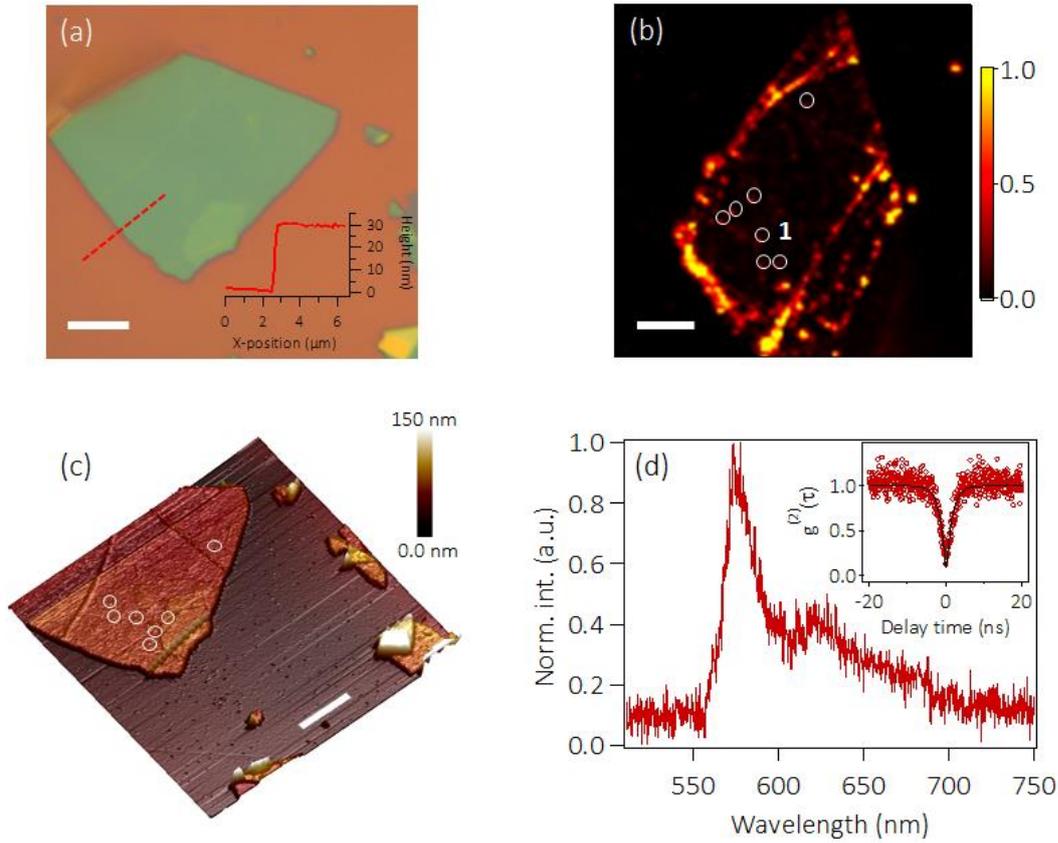

*Figure 2. Spatial characterisation of emitters in electron-irradiated, high purity multilayer flakes of single-crystal hBN. (a) Optical microscope image of a flake. The inset is an AFM line scan showing that the average thickness is ~30 nm. (b) Confocal PL scan showing emitters created by electron irradiation. The emitter positions are shown by white circles on the confocal map. (c) AFM image of the same flake showing the positions of the emitters in b (white circles). It is noted that all emitters shown here are on the flat region of the flake. (d) PL spectrum of emitter 1. The inset shows the corresponding $g^{(2)}(\square)$ autocorrelation function. All scale bars are 5 μm*

To further investigate the effects of high energy electron irradiation, we compared hBN samples grown in the presence of carbon (i.e. BN doped to a C concentration of ~$10^{19}$ cm$^{-3}$) and samples grown under standard conditions but using the $^{10}$B isotope. The isotope is not important for our experiments as we do not study spin or phononic effects. However, this sample serves as a useful reference material to the high purity hBN, since the atomic structures of the emitters and the roles of impurities are uncertain. PL spectra of emitters found in these flakes exhibit characteristics that are similar to the ones in the high purity hBN. Figure 3(a, d) show three PL spectra recorded from emitters in $^{10}$B and carbon-doped hBN, respectively. Most of the emitters have zero phonon line energies in the range of 580 - 600 nm, and a clear phonon sideband. The sharp edge around 565 nm is due to a bandpass filter used in our measurements. Figure 3 (b,e) confirm the antibunched nature of the emitters. Only one example is shown for clarity, however, most emitters found in the samples showed antibunching. Fig 3 (c, f) show the stability curves that confirm the emission is photostable and does not exhibit blinking or bleaching. Note that there was no annealing step involved in generating these emitters.

It has been suggested that carbon may play a role in these emitters. Our results do not support this assumption as we have not found any significant, systematic differences between the quantity (see Table 1) and optical characteristics of emitters in the carbon-doped material and the other flakes investigated in this study.

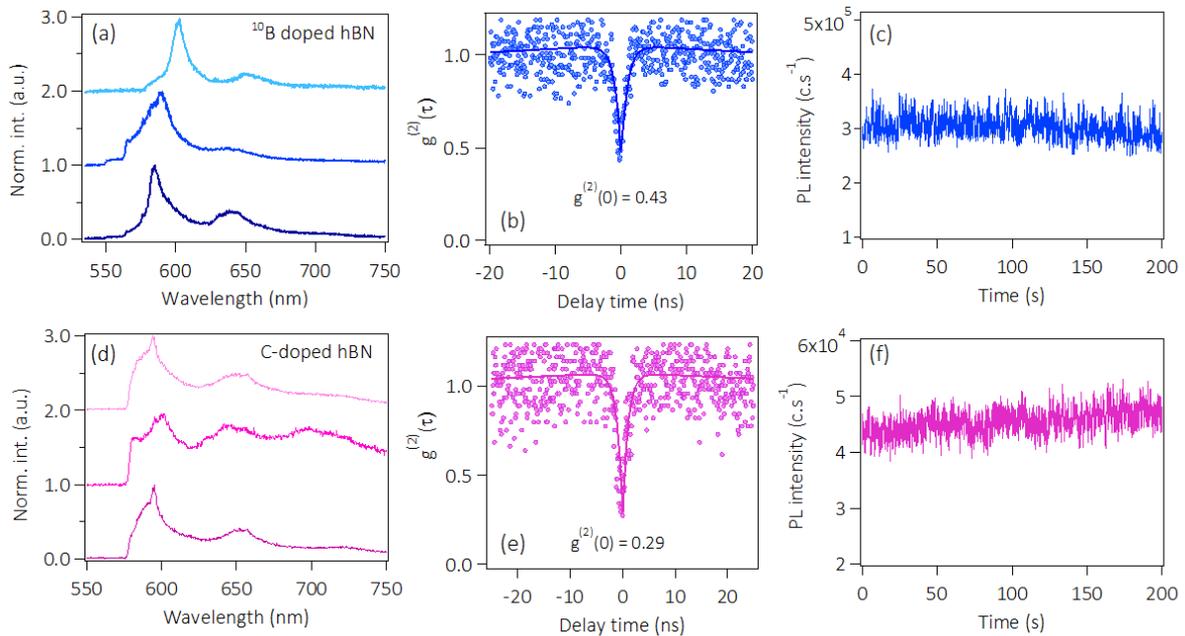

*Figure 3. Optical characterisation of emitters in electron-irradiated B10 and C-doped hBN samples. (a) Typical luminescence spectra of emitters in B10. (b) Second-order autocorrelation function of an emitter with $g^{(2)}(0) = 0.43$. A $g^{(2)}(0)$ value of less than 0.5 is indicative of single-photon emission (c) Corresponding stability curve of the aforementioned emitter. (d) Typical luminescence spectra of emitters in C-doped hBN. (e) An example of the second-order autocorrelation function of an emitter, with $g^{(2)}(0) = 0.29$. (f) Corresponding stability curve of the emitter. All optical characterisation was performed at room temperature with 532 nm CW excitation. All spectra were collected at 0.2 mW for 10 seconds.*

Cryogenic PL measurements were performed on emitters in electron-irradiated flakes to see whether the ZPLs approach spectrometer-limited resolution and have the potential to be used for advanced quantum optics experiments[32-33]. These PL measurements were taken at 10 K using a 707 nm excitation source. Figure 3a and 3b show two examples of ZPLs from the emitters, with a FWHM of 0.18 ± 0.05 nm and 0.35 ± 0.05 nm, respectively. Most of the studied emitters exhibit spectral fluctuations on the time scale of seconds, as shown in figures 3 (c, d), respectively. These results are consistent with emitters studied in samples that were not processed by electron irradiation and in commercially-available multilayers[17, 32].

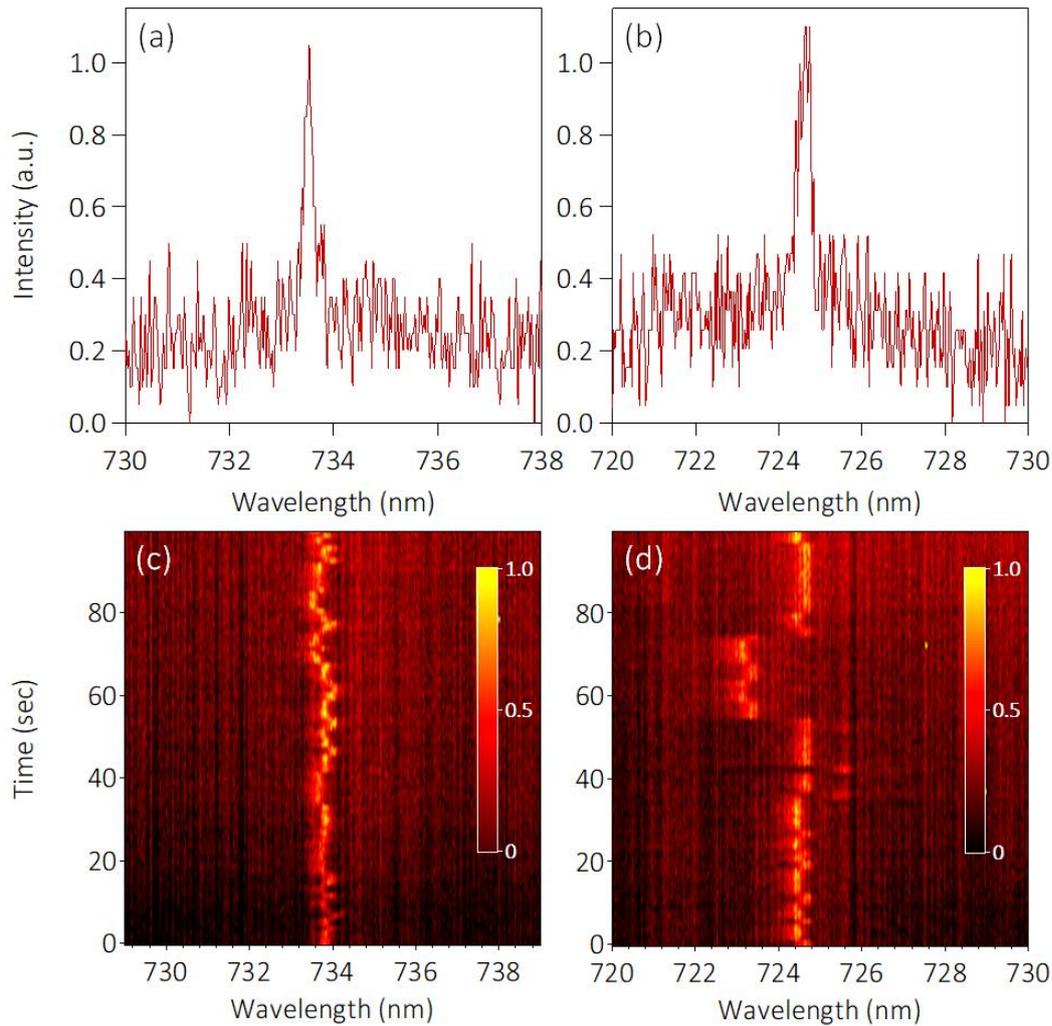

*Figure 4. Low temperature measurements of emitters found in electron-irradiated, high purity, single-crystal multilayer flakes. (a, b) Typical PL spectra recorded from two SPEs. (c, d) Spectra from the same emitters measured as a function of time. The emitters exhibit spectral diffusion. All measurements were performed at 10K.*

Finally, we studied CVD-grown monolayers of hBN. The monolayers were transferred from copper onto a silicon substrate (see methods). Figure 4a shows an atomic force microscope (AFM) image with a step of ~ 0.79 nm corresponding to a hBN monolayer. The white spots on the monolayer are PMMA residue from the transfer process. PL scans reveal a relatively high density of emitters after electron-irradiation (see Table 1). Figures 5 (b, c) show a typical example of a confocal map and a spectrum from a single emitter within the monolayers. The emitters have similar characteristics to those in large multi-layered flakes in terms of their emission wavelengths, FWHMs and phonon sidebands. Annealing for 1 hour at 750 °C under 1 torr of Ar results in the formation of new emitters within the monolayers. Figure 5(d, e) show a confocal map and PL from single emitters observed in an annealed monolayer, respectively. The annealing process destroyed the previously-characterised emitters, while simultaneously creating new ones. In general, emitters in monolayer hBN are less stable, more susceptible to being destroyed/deactivated by annealing, and exhibit more blinking and photobleaching under laser excitation than emitters in multilayer flakes of hBN - all likely due to a lack of encapsulation by the host material.

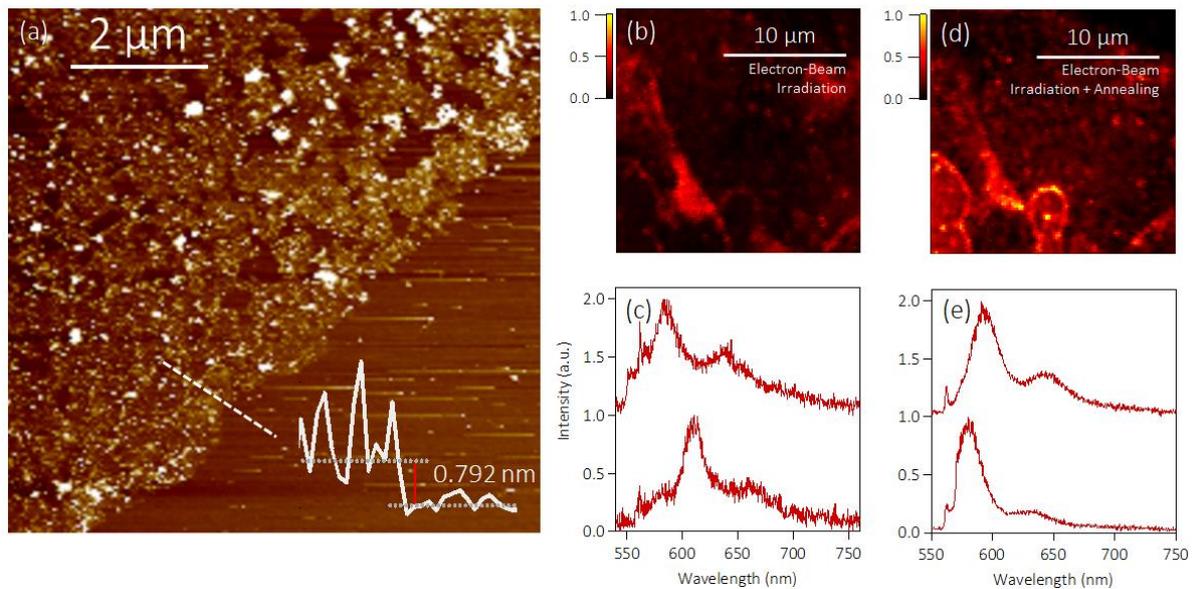

*Figure 5. Effects of electron irradiation on hBN monolayers. (a) AFM line scan showing an average monolayer thickness of 0.79 ± 0.40 nm. (b) PL scan of an area where a hBN monolayer was deposited. (c) Typical PL spectra of emitters formed by electron irradiation. (d) PL scan of the same region of the monolayer after high-temperature annealing. (e) Typical PL spectra of emitters formed after annealing of the irradiated samples.*

Table 1 summarizes the quantities of emitters found in all samples after MeV electron irradiation, and in control samples that were not irradiated but were instead subjected to the annealing treatment (1 hour at 750 °C under 1 torr of Ar). Clearly, electron irradiation is a promising method to create emitters without the need for annealing. The irradiation treatment consistently produced more emitters than the annealing process, and has the potential to generate emitters away from hBN flakes edges and grain boundaries. The samples that we investigated include C-enriched hBN, and our results do not support the hypothesis that C plays a role in the atomic structure of the emitters[26]. Another important observation is that only one 'class' of emitters was observed in all of our samples, all of which were either large flakes of mechanically-exfoliated multilayers, or large monolayers of CVD-grown hBN. Conversely, emitters in small, liquid-exfoliated flakes of hBN (supplied by Graphene Supermarket) can be broadly classified into two groups, the second of which is characterised by longer ZPL wavelengths and weak/negligible phonon sidebands. The absence of such emitters from all samples investigated in the present study suggests that they have a different atomic structure. On a practical level, the promising high energy electron irradiation technique presented here will likely accelerate understanding of the nature of single emitters in hBN, and their deployment in devices.

| **Statistics of emitter formation after MeV electron irradiation.** | | |
|---|---|---|
|  | *MeV irradiated* | *Annealed only* |
| *High purity hBN multilayers* | ~ 9 emitters per flake | 1 - 2 emitters per flake |
| *B-10 enriched hBN multilayers* | ~ 8 emitters per flake | 1 - 2 emitters per flake |
| *Carbon enriched hBN multilayers* | ~ 5 emitters per flake | 1 - 2 emitters per flake |
| *CVD-grown hBN monolayers* | ~ 5 emitters per area | ~ 3 emitters per area |

*Table 1. Distribution of emitters in various flakes. All flakes (and areas) were chosen to have a similar size of ~20 µm*

CONCLUSIONS

To conclude, we studied the effect of MeV electron irradiation on SPEs in hBN. Our results conclusively show that MeV irradiation can greatly enhance the formation of SPEs in hBN, without the necessity for a further annealing step. We also showed that upon irradiation, the emitters can form in the flat areas of the hBN flake. Finally, our work showed a similar distribution of emitters in carbon rich hBN and high purity hBN, therefore suggesting carbon may not be involved in the crystallographic structure of the defects. Our results are promising to develop a reliable procedure for engineering SPEs in hBN and unveiling their crystallographic structure.

**Acknowledgments**

Financial support from the Australian Research council (via DP140102721, DP180100077, LP170100150), the Asian Office of Aerospace Research and Development grant FA2386-17-1-4064, the Office of Naval Research Global under grant number N62909-18-1-2025 are gratefully acknowledged. K.W. and T.T. acknowledge support from the Elemental Strategy Initiative conducted by the MEXT, Japan and JSPS KAKENHI, Grant Number JP15K21722.